\documentclass[11pt]{article}
\usepackage{graphicx}
\usepackage{moriond}
\newcommand{\sss}{\scriptscriptstyle}
\begin{document}
\begin{flushright}
BNL-HET-03/10\\
FSU-HEP-2003-0513\\
UB-HET-03/04\\
hep-ph/0305282
\end{flushright}
\vspace*{3cm} 
\title{ASSOCIATED HIGGS BOSON PRODUCTION WITH HEAVY QUARKS} 
\author{S.~DAWSON$^1$, L.~H.~ORR$^2$, L.~REINA$^{3,\,}$\footnote{Talk
    presented by L.~Reina. at the XXXVIIIth Rencontres de Moriond,
    Les Arcs, March 15-22 2003}, D.~WACKEROTH$^4$}
\address{$^1$Physics Department, Brookhaven National Laboratory, Upton, NY
  11973, USA\\
$^2$Department of Physics and Astronomy, University of Rochester,
  Rochester, NY 14627, USA\\
$^3$Physics Department, Florida State University, Tallahassee, FL
  32306-4350, USA\\
$^4$Department of Physics, SUNY at Buffalo, Buffalo, NY 14260, USA}
\maketitle 
\abstracts{The production of a Higgs boson in association with a pair
  of $t\bar{t}$ quarks will play a very important role at both hadron
  and lepton colliders. We review the status of theoretical
  predictions and their relevance to Higgs boson studies, with
  particular emphasis on the recently calculated NLO QCD corrections
  to the inclusive cross section for $p\bar{p},pp\to t\bar{t}h$. We
  conclude by briefly discussing the case of exclusive $b\bar{b}h$
  production and the potential of this process in revealing signals of
  new physics beyond the Standard Model.}
\section{Introduction}
\label{sec:intro}

Present and future colliders will play a crucial role in exploring the
nature of the electroweak symmetry breaking and its relation to the
origin of fermion masses. The discovery of a Higgs boson is therefore
among the most important goals of both the Tevatron and the Large
Hadron Collider (LHC). Had such a particle to be discovered, a high
energy Linear Collider (LC) will be able to identify it unambiguously.

The present lower bounds on the Higgs boson mass from direct searches
at LEP2 are $M_h\!>\!114.4$~GeV (at $95\%$
CL)~\cite{Lephwg1:2001_2002} for the Standard Model (SM) Higgs boson
($h$), and $M_{h^0}\!>\!91.0$~GeV and $M_{A^0}\!>\!91.9$~GeV (at
$95\%$ CL, $0.5\!<\!\tan\beta\!<\!2.4$
excluded)~\cite{Lephwg1:2001_2002} for the light scalar ($h^0$) and
pseudoscalar ($A^0$) Higgs bosons of the minimal supersymmetric
standard model (MSSM).  At the same time, global SM fits to
electroweak precision data imply $M_{h}<211$~GeV (at
$95\%$~CL)~\cite{Lepewwg:2003}, while the MSSM requires the existence
of a scalar Higgs boson lighter than about 130~GeV. The possibility of
a Higgs boson discovery in the mass range near 115-130~GeV thus seems
increasingly likely.

In this context the associated production of a Higgs boson with a pair
of $t\bar{t}$ quarks is kinematically accessible, has a very
distinctive signature, and can give the only handle on a direct
measurement of the top quark Yukawa coupling, perhaps the most crucial
coupling in exploring the origin of fermion masses.  

Observing $p\bar{p}\to t\bar{t}h$ at the Tevatron
($\sqrt{s}\!=\!2$~TeV) will require very high
luminosity~\cite{Goldstein:2000bp} and will probably be at the edge of
the machine capabilities. On the other hand, if $M_{h}\!\le\!130$~GeV,
$pp\to t\bar{t}h$ is an important discovery channel for a SM-like
Higgs boson at the LHC
($\sqrt{s}\!=\!14$~TeV)~\cite{atlas:1999,Richter-Was:1999sa,Beneke:2000hk,Drollinger:2001ym}.
Given the statistics expected at the LHC, $pp\to t\bar{t}h$, with
$h\to b\bar{b},\tau^+\tau^-,W^+W^-,\gamma\gamma$ will also be
instrumental to the determination of the couplings of a discovered
Higgs
boson~\cite{Beneke:2000hk,Zeppenfeld:2000td,Belyaev:2002ua,Maltoni:2002jr}.
Several analyses show that precisions of the order of 10-15\% on the
measurement of the top quark Yukawa coupling can be obtained with
integrated luminosities of 100~fb$^{-1}$ per detector.  Morever, the
combined measurements of $pp\to t\bar{t}h$ with $h\to b\bar{b}$ and
$h\to\tau^+\tau^-$ could provide the only model independent
determination of the ratio of the bottom quark to the $\tau$ lepton
Yukawa couplings~\cite{Belyaev:2002ua}.

At a LC, the top quark Yukawa coupling can be measured in a model
independent way via $e^+e^-\to t\bar{t}h$.  The inclusive cross
section for $e^+e^-\to t\bar{t}h$ (and $b\bar{b}h$) has been
calculated including the first order of QCD corrections, both in the
SM and in the MSSM~\cite{Dittmaier:1998dz,Dawson:1998ej}, and the
theoretical uncertainty is reduced in this case to less than 10\%.
However, the precision of the measurement is severely limited by the
machine center of mass energy.  Dedicated studies show
that~\cite{Juste:1999af}, at the optimal center of mass energy of
$\sqrt{s}\!\simeq\!800$~GeV, integrated luminosities of 1000~fb$^{-1}$
will allow to determine the top quark Yukawa coupling at the 5\%
level, for $M_h\!=\!120$~GeV.  However, at a center of mass energy of
$\sqrt{s}\!=\!500$~GeV the $e^+e^-\to t\bar{t}h$ event rate is tiny
and, for the same range of Higgs masses and integrated luminosity, a
LC will initially measure the top Yukawa coupling with precisions of
at best 20\%~\cite{Juste:1999af,Baer:1999ge}. Given this intrinsic
limitation, the role played by the LHC and, in this context, by the
associated production of a Higgs boson with a pair of $t\bar{t}$
quarks becomes even more important.

In view of its phenomenological relevance, a lot of effort has been
recently invested in improving the stability of the theoretical
predictions for the hadronic inclusive total cross section for
$p\bar{p},pp\to t\bar{t}h$. Since the tree level or Leading Order (LO)
cross section is affected by a very large renormalization and
factorization scale dependence, the first order of QCD corrections
have been calculated and the Next-to-Leading (NLO) cross section, for
a SM Higgs boson, has been obtained independently by two
groups~\cite{Beenakker:2001rj,Reina:2001sf,Dawson:2002tg}.  The NLO
cross section has a drastically reduced renormalization and
factorization scale dependence, of the order of 15\% as opposed to the
initial 100\% uncertainty of the LO cross section, and leads to
increased confidence in predictions based on these results.

The calculation of the NLO corrections to the hadronic process
$p\bar{p},pp\to t\bar{t}h$ presents challenging technical
difficulties, ranging from virtual pentagon diagrams with several
massive internal and external particles to real gluon and quark
emission in the presence of infrared singularities. A general overview
of the techniques developed and employed in our calculation are
presented in Section~\ref{sec:calculation}, and the corresponding
results are illustrated in Section~\ref{sec:results}. We conclude with
a brief outlook in Section~\ref{sec:outlook}.
\boldmath
\section{QCD corrections to $t\bar{t}h$ production at the Tevatron and
  the LHC}
\label{sec:calculation}
\unboldmath
The inclusive total cross section for $pp\to t\bar{t}h$ at ${\cal
  O}(\alpha_s^3)$ can be written as:
\begin{equation}
\label{eq:sigma_nlo}
\sigma_{\sss NLO}(p\,p\hskip-7pt\hbox{$^{^{(\!-\!)}}$}\to t\bar{t}h)=
\sum_{ij}\frac{1}{1+\delta_{ij}}
\int dx_1 dx_2 \left[{\cal F}_i^p(x_1,\mu) {\cal F}_j^{p(\bar{p})}(x_2,\mu)
{\hat \sigma}^{ij}_{\sss NLO}(x_1,x_2,\mu)+(1\leftrightarrow 2)\right]
\,\,\,,
\end{equation}
where ${\cal F}_i^{p(\bar{p})}$ are the NLO parton distribution
  functions (PDFs) for parton $i$ in a (anti)proton, defined at a
  generic factorization scale $\mu_f\!=\!\mu$, and ${\hat
    \sigma}^{ij}_{\sss NLO}$ is the ${\cal O}(\alpha_s^3)$
  parton-level total cross section for incoming partons $i$ and $j$,
  made of the channels $q\bar{q},gg\to t\bar{t}h$ and $(q,\bar{q})g\to
  t\bar{t}h(q,\bar{q})$, and renormalized at an arbitrary scale
  $\mu_r$ which we also take to be $\mu_r\!=\!\mu$.  We note that the
  effect of varying the renormalization and factorization scales
  independently has been investigated and found to be negligible.  The
  partonic center of mass energy squared, $s$, is given in terms of
  the hadronic center of mass energy squared, $s_{\sss H}$, by $s=x_1
  x_2 s_{\sss H}$. At the Tevatron center of mass energy the cross
  section is entirely dominated by the $q\bar{q}$ initial state and
  the results presented in Section~\ref{sec:results} are obtained by
  including only $q\bar{q}\to t\bar{t}h$ at the parton level. At the
  LHC center of mass energy the cross section is dominated by the $gg$
  initial state, but the other contributions cannot be neglected and
  are included in our calculation.

We write the NLO parton-level total cross section ${\hat
  \sigma}_{\sss NLO}^{ij}(x_1,x_2,\mu)$ as:
\begin{equation}
\label{eq:sigmahat_nlo}
{\hat\sigma}_{\sss NLO}^{ij}(x_1,x_2,\mu)\equiv
{\hat \sigma}_{\sss LO}^{ij}(x_1,x_2,\mu)+
\delta {\hat \sigma}_{\sss NLO}^{ij}(x_1,x_2,\mu)\,\,\,,
\end{equation}
where ${\hat\sigma}_{\sss LO}^{ij}(x_1,x_2,\mu)$ is the ${\cal O}(\alpha_s^2)$ Born cross
section, and $\delta{\hat\sigma}_{\sss NLO}^{ij}(x_1,x_2,\mu)$
consists of the ${\cal O}(\alpha_s)$ corrections to the Born cross
sections for $gg,q\bar{q}\to t\bar{t}h$ and of the tree level
$(q,\bar{q})g\to t\bar{t}h(q,\bar{q})$ processes, including the
effects of mass factorization.
$\delta{\hat\sigma}_{\sss NLO}^{ij}(x_1,x_2,\mu)$ can be written as
the sum of two terms:
\begin{eqnarray}
\label{eq:delta_sigmahat}
\delta{\hat\sigma}_{\sss NLO}^{ij}(x_1,x_2,\mu)&=&
\int d(PS_3) \overline{\sum}|{\cal A}_{virt}(ij\to t\bar{t}h)|^2+
\int d(PS_4)\overline{\sum}|{\cal A}_{real}(ij\to t\bar{t}h+l)|^2
\nonumber \\
&\equiv&\hat{\sigma}^{ij}_{virt}(x_1,x_2,\mu)+
\hat{\sigma}^{ij}_{real}(x_1,x_2,\mu)\,\,\,,
\end{eqnarray}
where $|{\cal A}_{virt}(ij\to t\bar{t}h)|^2$ and $|{\cal
  A}_{real}(ij\to t\bar{t}h+l)|^2$ (for $ij\!=\!q\bar{q},gg$ and
$l\!=\!g$, or $ij\!=\!qg,\bar{q}g$ and $l\!=\!q,\bar{q}$) are
respectively the ${\cal O}(\alpha_s^3)$ terms of the squared matrix
elements for the $ij\rightarrow t\bar th$ and $ij\rightarrow t\bar t
h+l$ processes, and $\overline{\sum}$ indicates that they have been
averaged over the initial state degrees of freedom and summed over the
final state ones.  Moreover, $d(PS_3)$ and $d(PS_4)$ in
Eq.~(\ref{eq:delta_sigmahat}) denote the integration over the
corresponding three and four-particle phase spaces respectively.  The
first term in Eq.~(\ref{eq:delta_sigmahat}) represents the
contribution of the virtual one gluon corrections to $q\bar{q}\to
t\bar{t}h$ and $gg\to t\bar{t}h$, while the second one is due to the
real one gluon and real one quark/antiquark emission, i.e.
$q\bar{q},gg\to t\bar{t}h+g$ and $qg(\bar qg) \to
t\bar{t}h+q(\bar{q})$.

The ${\cal O}(\alpha_s)$ virtual and real corrections to $q\bar{q}\to
t\bar{t}h$ and $gg\to t\bar{t}h$ have been discussed in detail in
Refs.~\cite{Reina:2001sf,Dawson:2002tg} and we will highlight in the
following only the most challenging tasks. 
\subsection{Virtual correction}
\label{subsec:sigma_virtual}

The calculation of the ${\cal O}(\alpha_s)$ virtual corrections to
$q\bar{q},gg\to t\bar{t}h$ proceeds by reducing each virtual diagram
to a linear combination of tensor and scalar integrals, which may
contain both ultraviolet (UV) and infrared (IR) divergences.  Tensor
integrals are further reduced in terms of scalar
integrals~\cite{Passarino:1979jh}.  The finite scalar integrals are
evaluated by using the method described in Ref.~\cite{Denner:1993kt}
and cross checked with the FF package~\cite{vanOldenborgh:1990wn}.
The scalar integrals that exhibit UV and/or IR divergences are
calculated analytically. Both the UV and IR divergences are extracted
by using dimensional regularization in $d\!=\!4-2\epsilon$ dimensions.
The UV divergences are then removed by introducing a suitable set of
counterterms, as described in detail in
Refs.~\cite{Reina:2001sf,Dawson:2002tg}. The remaining IR divergences
are cancelled by the analogous singularities in the soft and collinear
part of the real gluon emission cross section.

The most difficult integrals arise from the IR-divergent pentagon
diagrams with several massive particles.  The pentagon scalar and
tensor Feynman integrals originating from these diagrams present
either analytical (scalar) or numerical (tensor) challenges. We have
calculated the pentagon scalar integrals as linear combinations of
scalar box integrals using the method of Ref.~\cite{Bern:1993em}, and
cross checked them using the techniques of Ref.~\cite{Denner:1993kt}.
Pentagon tensor integrals can give rise to numerical instabilities due
to the dependence on inverse powers of the Gram determinant (GD),
GD$\!=\!\det(p_i\!\cdot\!p_j)$ for $p_i$ and $p_j$ external momenta,
which vanishes at the boundaries of phase space when two momenta
become degenerate. These are spurious divergences, which cause serious
numerical difficulties.  To overcome this problem we have calculated
and cross checked the pentagon tensor integrals in two ways:
numerically, by isolating the numerical instabilities and
extrapolating from the numerically safe to the numerically unsafe
region using various techniques; and analytically, by reducing them to
a numerically stable form.
\subsection{Real correction}
\label{subsec:sigma_real}
In computing the ${\cal O}(\alpha_s)$ real corrections to
$q\bar{q},gg\to t\bar{t}h$ and $(q,\bar{q}g\to
t\bar{t}h+(q,\bar{q})$ it is crucial to isolate the IR divergent
regions of phase space and extract the corresponding singularities
analytically.  We achieve this by using the phase space slicing (PSS)
method, in both the double~\cite{Harris:2001sx} and
single~\cite{Giele:1992vf,Keller:1998tf} cutoff approaches. In both
approaches the IR region of the $t\bar{t}h+g$ phase space where the
emitted gluon cannot be resolved is defined as the region where the
gluon kinematic invariants:
\begin{equation}
s_{ig}= 2 p_i\cdot p_g=2E_iE_g(1-\beta_i\cos\theta_{ig})
\end{equation}
become small. Here $p_i$ is the momentum of an external (anti)quark or
gluon (with energy $E_i$), $\beta_i\!=\!\sqrt{1-m_i^2/E_i^2}$, $p_g$
is the momentum of the radiated final state gluon ((anti)quark) (with
energy $E_g$), and $\theta_{ig}$ is the angle between $\vec{p}_i$ and
$\vec{p}_g$.  In the IR region the cross section is calculated
analytically and the resulting IR divergences, both soft and
collinear, are cancelled, after mass factorization, against the
corresponding divergences from the ${\cal O}(\alpha_s)$ virtual
corrections.

The single cutoff PSS technique defines the IR region as that where
\begin{equation}
s_{ig}<s_{min}\,\,\,,
\end{equation}
for an arbitrarily small cutoff $s_{min}$.  The two cut-off PSS method
introduces two arbitrary parameters, $\delta_s$ and $\delta_c$, to
separately define the IR soft and IR collinear regions according to:
\begin{eqnarray}
&&E_g<{\delta_s\sqrt{s}\over 2}\,\,\,\,\,\mbox{soft region}\,\,\,,\nonumber\\
&&(1-\cos\theta_{ig})<\delta_c\,\,\,\,\,\mbox{collinear region}\,\,\,.
\end{eqnarray}

In both methods, the real contribution to the NLO cross section is
computed analytically below the cutoffs and numerically above the
cutoffs, and the final result is independent of these arbitrary
parameters. With this respect, it is crucial to study the behavior of
$\sigma_{\sss NLO}$ in a region where the cutoff(s) are small enough
to justify the analytical calculations of the IR divergent
contributions to the real cross section, but not so small as to cause
numerical instabilities.
\boldmath
\section{Results for $t\bar{t}h$ production at hadron colliders}
\label{sec:results}
\begin{figure}[ht]
\begin{minipage}{0.46\linewidth}
\centering
\includegraphics[scale=0.42]{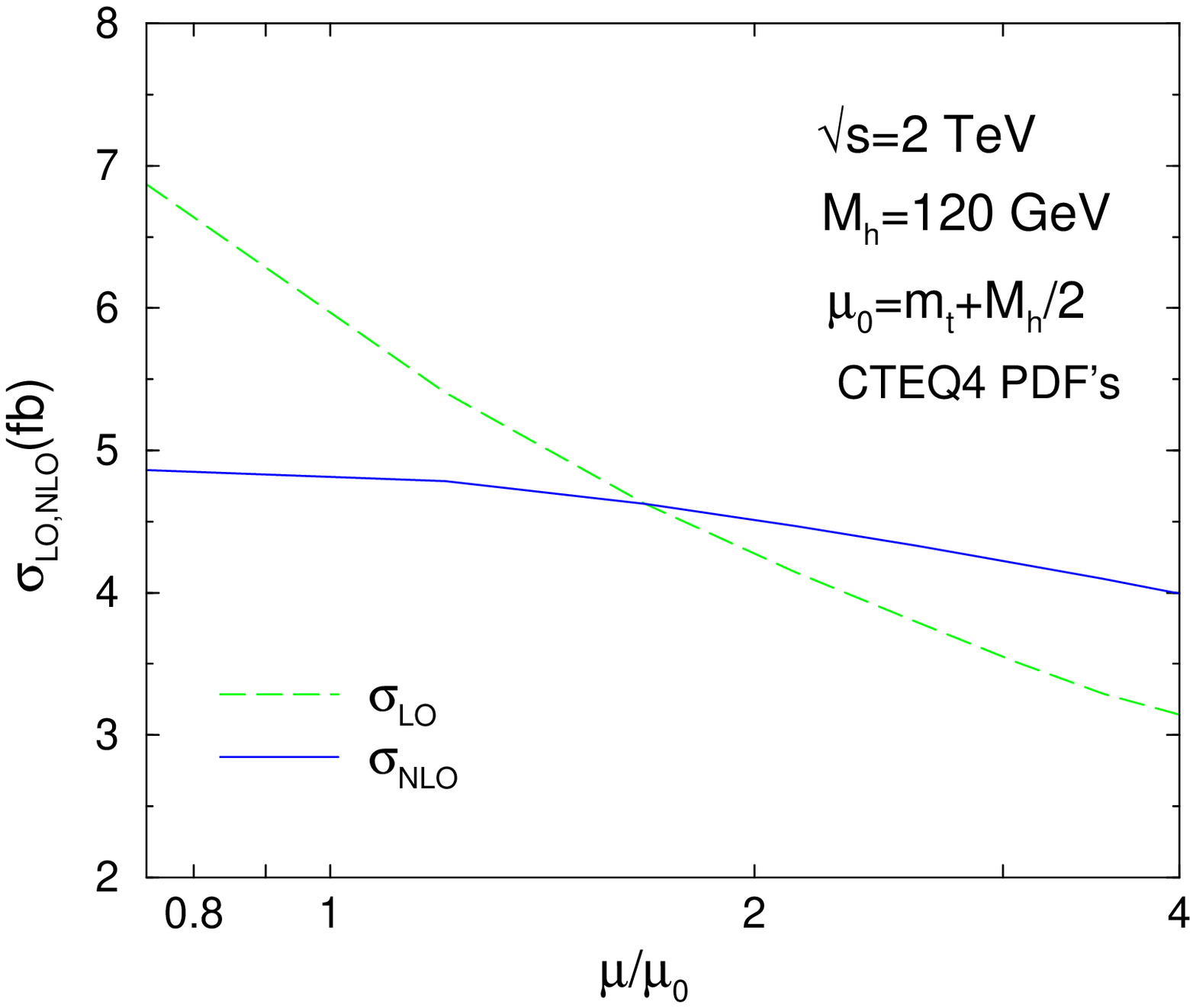}
\caption[]{Dependence of $\sigma_{\sss LO,NLO}(p\bar{p}\to t\bar{t}h)$ on 
  the renormalization/factorization scale $\mu$, at $\sqrt{s_{\sss
      H}}\!=\!2$~TeV, for $M_h\!=\!120$ GeV.}
\label{fg:mu_dependence_tev}
\end{minipage}
\hfill
\begin{minipage}{0.46\linewidth}
\centering
\includegraphics[scale=0.42]{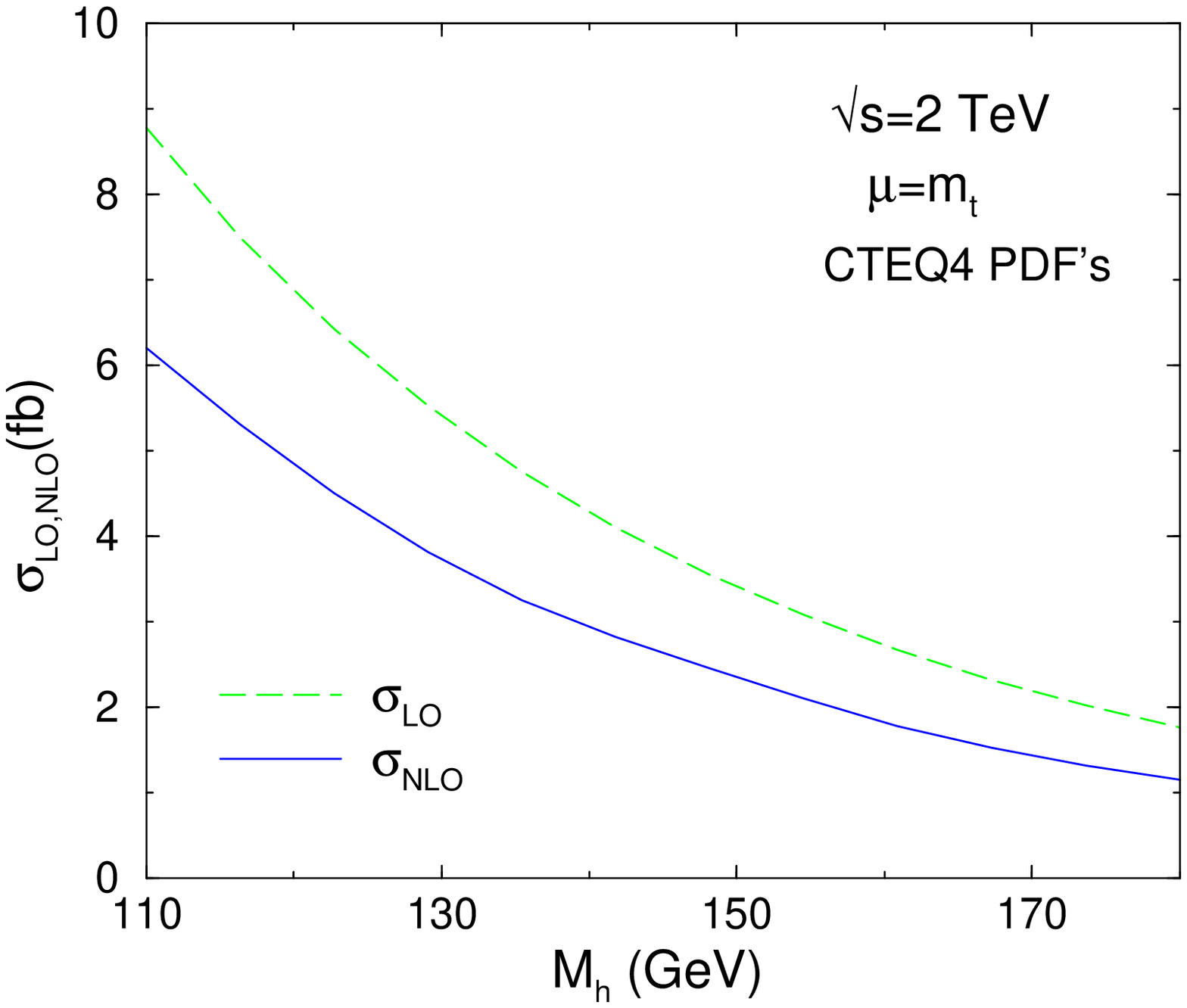}
\caption[]{$\sigma_{\sss NLO}(p\bar{p}\to t\bar{t}h)$ and $\sigma_{\sss
    LO}(pp\to t\bar{t}h)$ as functions of $M_h$, at $\sqrt{s_{\sss
      H}}\!=\!2$~TeV, for $\mu\!=m_t$.}
\label{fg:mh_dependence_tev}
\end{minipage}
\end{figure}

\begin{figure}[ht]
\begin{minipage}{0.46\linewidth}
\centering
\includegraphics[scale=0.42]{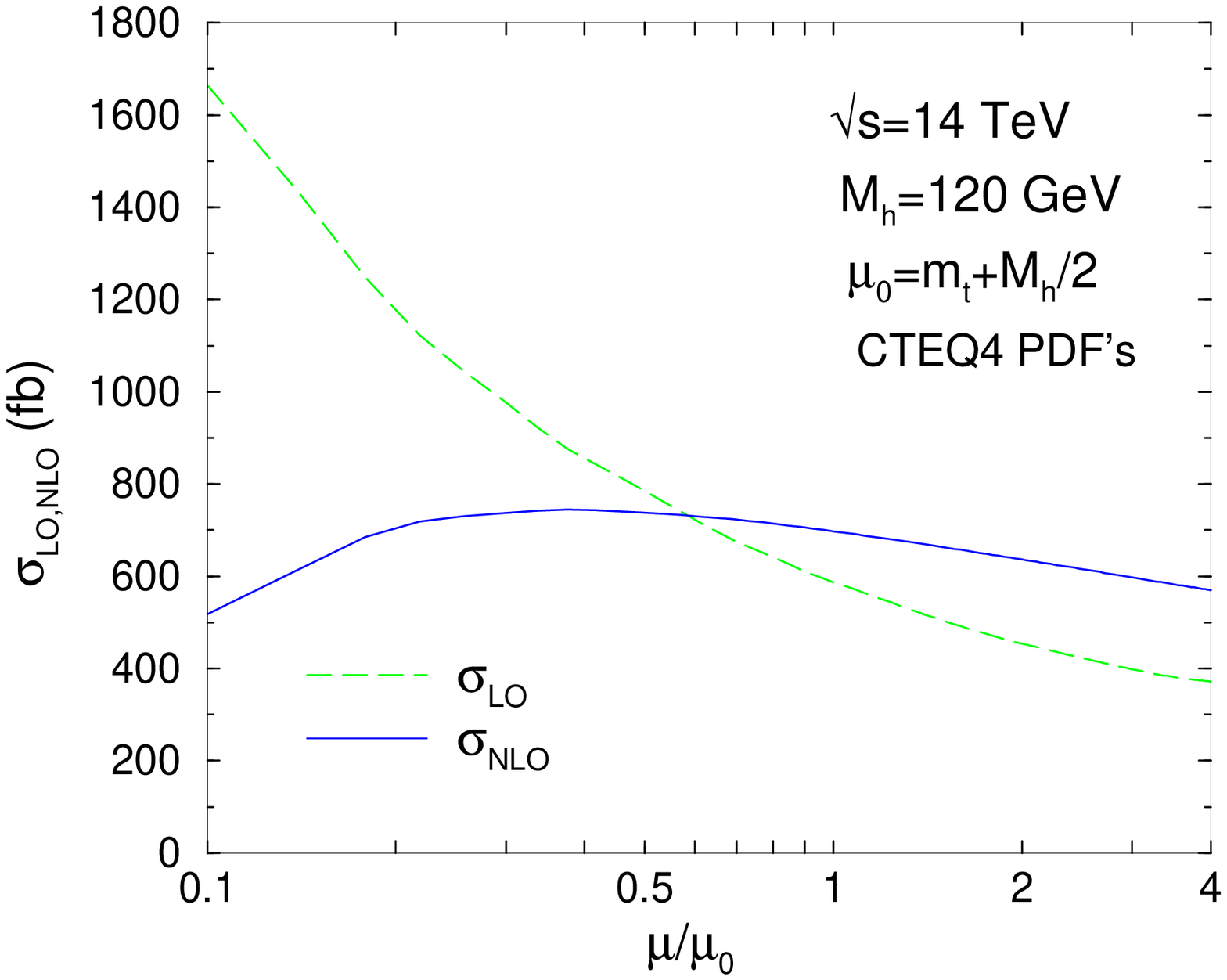}
\caption[]{Dependence of $\sigma_{\sss LO,NLO}(pp\to t\bar{t}h)$ on 
  the renormalization/factorization scale $\mu$, at $\sqrt{s_{\sss
      H}}\!=\!14$~TeV, for $M_h\!=\!120$ GeV.}
\label{fg:mu_dependence_lhc}
\end{minipage}
\hfill
\begin{minipage}{0.46\linewidth}
\centering
\includegraphics[scale=0.42]{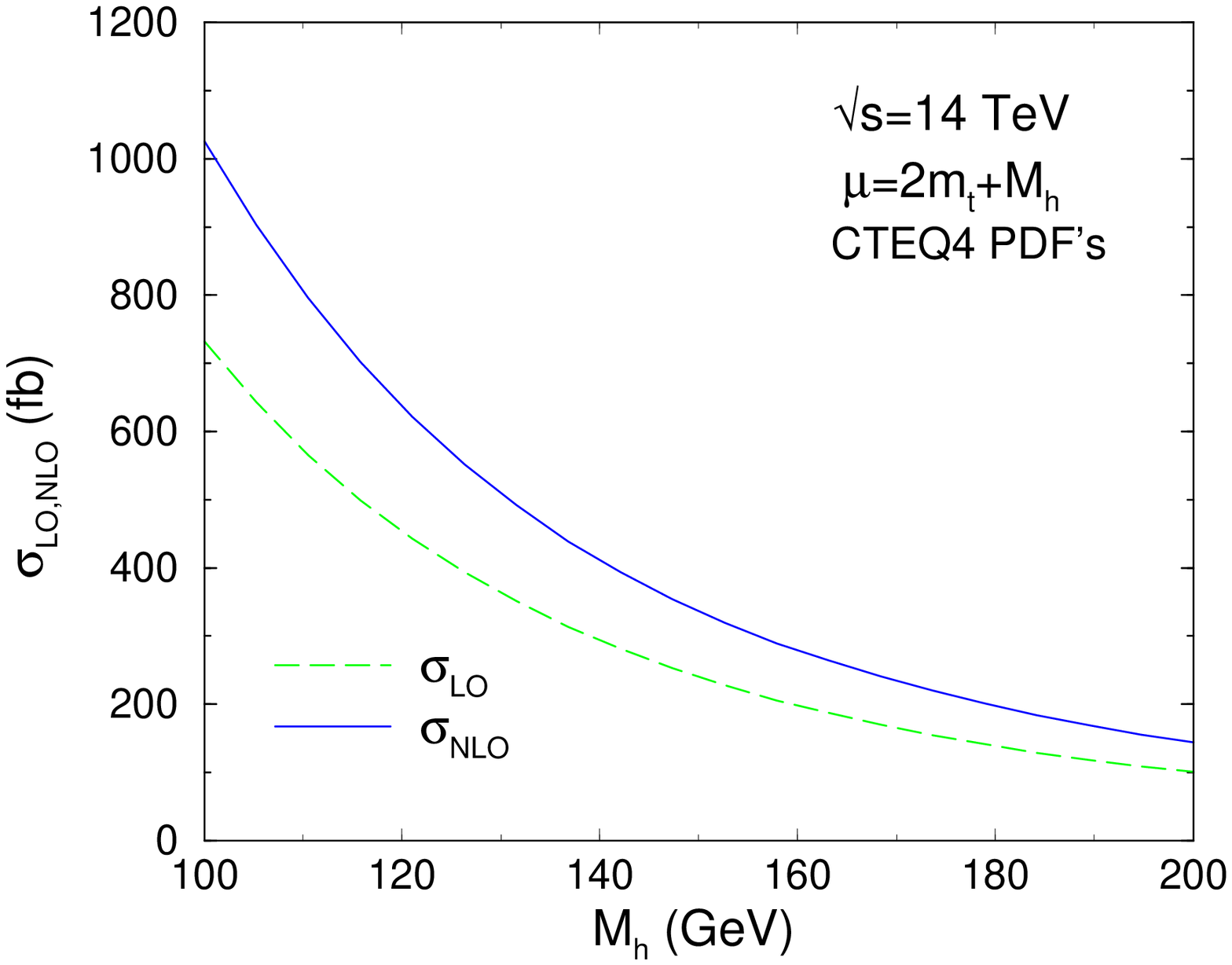}
\caption[]{$\sigma_{\sss NLO}(pp\to t\bar{t}h)$ and $\sigma_{\sss
    LO}(pp\to t\bar{t}h)$ as functions of $M_h$, at $\sqrt{s_{\sss
      H}}\!=\!14$~TeV, for $\mu\!=2m_t+M_h$.}
\label{fg:mh_dependence_lhc}
\end{minipage}
\end{figure}

\unboldmath
The impact of NLO QCD corrections on the tree level cross section is
summarized in
Figs.~\ref{fg:mu_dependence_tev}-\ref{fg:mh_dependence_lhc} for both
the Tevatron and the LHC. Results for $\sigma_{\sss LO}$ are obtained
using the 1-loop evolution of $\alpha_s(\mu)$ and CTEQ4L parton
distribution functions \cite{Lai:1997mg}, while results for
$\sigma_{\sss NLO}$ are obtained using the 2-loop evolution of
$\alpha_s(\mu)$ and CTEQ4M parton distribution functions, with
$\alpha_s^{\sss NLO}(M_Z)\!=\!0.116$.

Figs.~\ref{fg:mu_dependence_tev} and \ref{fg:mu_dependence_lhc}
illustrate the renormalization/factorization scale dependence of
$\sigma_{\sss LO}$ and $\sigma_{\sss NLO}$ at the Tevatron and the
LHC. In both cases the NLO cross section shows a drastic reduction of
the scale dependence with respect to the lowest order prediction.
Figs.~\ref{fg:mh_dependence_tev} and \ref{fg:mh_dependence_lhc}
complement this information by illustrating the dependence of the LO
and NLO cross sections on the Higgs boson mass at both the Tevatron
and the LHC.

The overall uncertainty on the theoretical prediction, including the
errors coming from parton distribution functions and the top quark
mass (which we take to be $m_t\!=\!174$~GeV), is reduced to only
15-20\%, as opposed to the 100-200\% uncertainty of the LO cross
section. Including NLO QCD corrections decreases (Tevatron) or
increases (LHC) the LO cross section for a broad range of commonly
used renormalization and factorization scales (obtained \emph{e.g.} by
varying $\mu$ by a factor of two around $\mu\!=\!\mu_0$), and over the
entire Higgs boson mass range considered in our study. This can be
summarized by defining a $K$-factor, $K=\sigma_{\sss NLO}/\sigma_{\sss
  LO}$, which is however affected by the same strong scale dependence
as the LO cross section, as well as by the choice of PDFs. When using
CTEQ4 PDFs the K-factor corresponding to
Figs.~\ref{fg:mu_dependence_tev}-\ref{fg:mh_dependence_lhc} is around
$0.7\!-\!0.95$ at the Tevatron and $1.2\!-\!1.4$ at the LHC, for most
choices of scales and Higgs boson mass.
\section{Outlook}
\label{sec:outlook}

The techniques developed to calculate the NLO cross section for
$p\bar{p},pp\to t\bar{t}h$ can now be applied to the study of the
associated production of $b\bar{b}h$. The inclusive cross section for
$b\bar{b}h$ production receives contributions from $b\bar{b}\to h$,
$bg\to bh$, and $gg\to b\bar{b}h$, in order of decreasing cross
section ($q\bar{q}\to b\bar{b}h$ is negligible at both the Tevatron
and the LHC). On the other hand, the exclusive cross section, corresponding to
the experimental situation when both final state $b$ quarks are
tagged, receives contributions from $gg\to b\bar{b}h$ only and can be
directly calculated, including NLO corrections, from the corresponding
results for $gg\to t\bar{t}h$. In spite of the smaller cross section,
the exclusive process is experimentally very interesting since it
corresponds to a well defined maesurement, where final state $b$ jets
are isolated via cuts on the transverse momentum of the $b$ and
$\bar{b}$ quarks. The cross section for $gg\to b\bar{b}h$ is negligible
in the SM and the detection of a Higgs boson in this channel would
unambiguously signal the presence of new physics responsible for an
anomalously large bottom quark Yukawa coupling, like the MSSM. This
could actually be a unique opportunity within the kinematical reach of
the Tevatron.

\section*{Acknowledgments}
The work of S.D. (L.H.O., L.R.) is supported in part by the U.S.
Department of Energy under grant DE-AC02-76CH00016
(DE-FG-02-91ER40685, DE-FG02-97ER41022). L.R. thanks the National
Science Foundation for supporting her participation to this meeting.
\section*{References}
\bibliography{/u/reina/tth/full/pp/qcd/gg/papers/long/laura/tth_lhc}
\end{document}